\documentclass[submission, PhysProc]{SciPost}

\hypersetup{
    colorlinks,
    linkcolor={red!50!black},
    citecolor={blue!50!black},
    urlcolor={blue!80!black}
}

\usepackage[bitstream-charter]{mathdesign}
\DeclareSymbolFont{usualmathcal}{OMS}{cmsy}{m}{n}
\DeclareSymbolFontAlphabet{\mathcal}{usualmathcal}

\newcommand\mf{\mathfrak}

\begin{document}

\begin{center}{\Large \textbf{
Derivation of Relativistic
Yakubovsky
Equations \\
under Poincar\'e Invariance
\\
}}\end{center}

% TODO: write the author list here. Use first name (+ other initials) + surname format.
% Separate subsequent authors by a comma, omit comma and use "and" for the last author.
% Mark the corresponding author with a superscript star.
\begin{center}
Hiroyuki Kamada \textsuperscript{$\star$}%,
%Aah B. Cee\textsuperscript{2} and
%Gee K. See\textsuperscript{3$\star$}
\end{center}

\begin{center}
%{\bf 1} 
Department of Physics, Faculty of Engineering, Kyushu Institute of Technology, \\
Kitakyushu 804-8550, Japan
%\\
%{\bf 2} Affiliation2
%\\
%{\bf 3} Affiliation2
\\
${}^\star$ {\small \sf kamada@mns.kyutech.ac.jp}
\end{center}

\begin{center}
\today
\end{center}

% For convenience during refereeing (optional),
% you can turn on line numbers by uncommenting the next line:
%\linenumbers

\definecolor{palegray}{gray}{0.95}
\begin{center}
\colorbox{palegray}{
  \begin{tabular}{rr}
  \begin{minipage}{0.05\textwidth}
    \includegraphics[width=14mm]{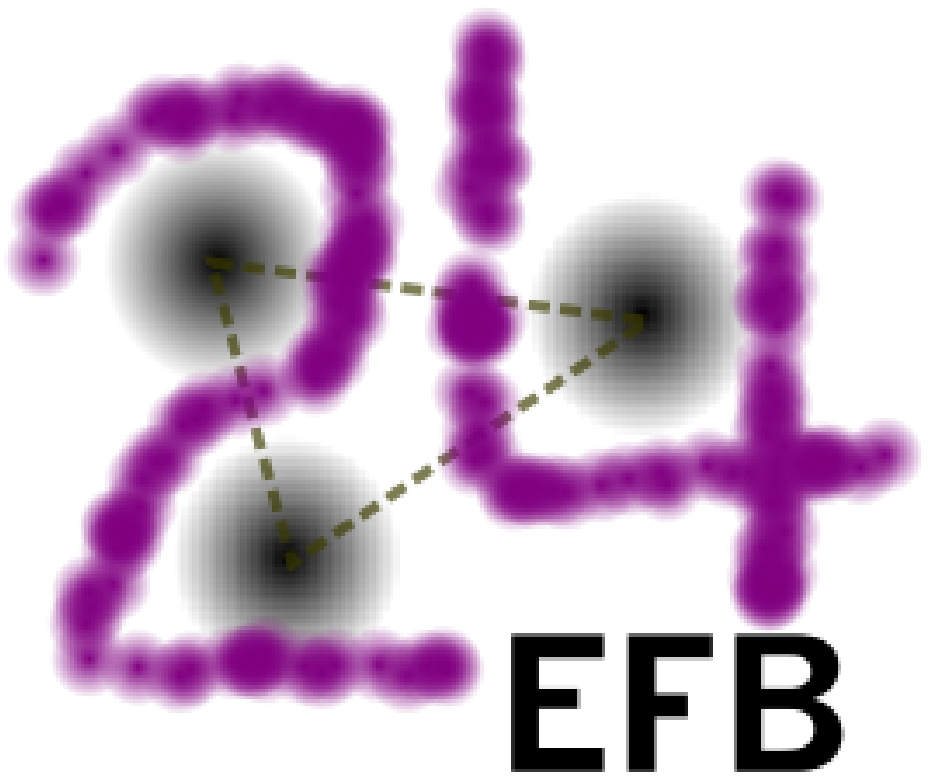}
  \end{minipage}
  &
  \begin{minipage}{0.82\textwidth}
    \begin{center}
    {\it Proceedings for the 24th edition of European Few Body Conference,}\\
    {\it Surrey, UK, 2-4 September 2019} \\
    %\doi{10.21468/SciPostPhysProc.2}\\
    \end{center}
  \end{minipage}
\end{tabular}
}
\end{center}

\section*{Abstract}
{\bf
% TODO: write your abstract here.
Relativistic Faddeev-Yakubovsky four-nucleon scattering equations are derived including a 
3-body force. We present these equations in the momentum space representation.
The quadratic integral equations using the iteration method, in order to obtain boosted potentials and 3-body force, are demonstrated.
}

\vspace{10pt}
\noindent\rule{\textwidth}{1pt}
\tableofcontents\thispagestyle{fancy}
\noindent\rule{\textwidth}{1pt}
\vspace{10pt}

\section{Introduction}
\label{sec:intro}
% TODO: write your article here.

At high energies one could expect deficiencies in the nonrelativistic Faddeev approach~\cite{Faddeev:1960su,Gloeckle:1995jg} in three-nucleon system. We have been constructing a relativistic framework 
in the form of relativistic Faddeev equations ~\cite{Glockle:1986zz,Keister:1991sb,Kamada:1999wy,Kamada:1999fz,Kamada:2007ms,Witala:2011yq,Polyzou:2010kx,Kamada:2014dba} according to the Bakamjian-Thomas 
theory~\cite{Bakamjian:1953kh}. 
Not only using the realistic nonrelativistic nucleon-nucleon (NN) potentials but using  
Kharkov relativistic NN potential \cite{Arslanaliev:2018wkl}  we obtained the triton wave function 
by solving the relativistic Faddeev equation~\cite{Kamada:2003fh,Kamada:2008xc,Kamada:2017mmk}.   
However, in the three-body scattering states, the relativistic effects appear to be generally small 
\cite{Sekiguchi:2005vq,Maeda:2007zza}
and insufficient to significantly improve the data description.
For sensitive observations such as $A_y$ puzzles, the relativistic effect has certainly 
surfaced~\cite{Witala:2008va}, but the results obtained were in the direction of deterioration.

As the number of particles handled increases, subtle  relativistic effects will be accumulated and surface, so here we would like to rewrite the Yakubovsky equations~\cite{Yakubovsky:1966ue}, which solve the four-nucleon system exactly, to a relativistic equations as well.

In Section 1 we organize relativistic momenta and its Jacobian. Section 2 deals with rewriting interaction to relativistic potential by Lorentz boost. The boosted potential satisfies the relativistic Lippmann-Schwinger (LS) equation in Sec 3. 
Section 4 looks back on how the 3-body Faddeev equation was relativistically transformed using the boosted potential. In Section 5, we will remodel the four-body Yakubovsky equation and derive the relativistic Yakubovsky equation. In section 6, we derive relativistic  equations involving 3-body force.  A summary is in section 7.
\section{2-body center of mass system}
\label{2-body}

In 2-body systems, these static mass of the particle are given  $m_i$ ($i = 1, 2$).
The four dimensional intrinsic momenta $\mf{p}_i$ are
\begin{eqnarray}
\mf{p}_i = (p^\mu_i)=(p_i^0, \vec p_i)=(E_i(p_i), p_{i}^x, p_{i}^y, p_{i}^z)
\end{eqnarray}
with 
\begin{eqnarray}
E_i(p_i) =\sqrt{m_i^2 + p_i^2}=\sqrt{m_i^2+\vec p_i \cdot \vec p_i}
\end{eqnarray}
By Lorentz transformation $L=(L_\mu ^\nu)$ of boosting velocity $\vec \upsilon$ 
the transformed four-momenta $\overline{ \mf{p}_i}$ are obtained
\begin{eqnarray}
\overline{ \mf{p}_i }=L \mf{p}_i =(\overline{ E_i} ,~\overline{ \vec p_i})=
(\gamma (E_i -\vec p_i\cdot \vec \upsilon),~\vec p_i +(\gamma-1)(\vec p_i \cdot \hat \upsilon ) \hat \upsilon  -\gamma E_i \vec \upsilon ).
\end{eqnarray}
with 
\begin{eqnarray}
\gamma \equiv {1\over \sqrt{1-\upsilon^2}}
\end{eqnarray}
where $\hat \upsilon$ is the unit vector of $\vec \upsilon$ and light velocity is set to 1. 

The relativistic total momentum $\mf{P}_{12}$ and the relative momentum $\vec k_{12}$ are given
~\cite{Fong:1986xx} as
\begin{eqnarray}
\mf{P}_{12} \equiv \mf{p}_1 + \mf{p}_2 =(E_1+E_2, \vec p_1+\vec p_2)=(E_{12},\vec P_{12}),
\end{eqnarray}
\begin{eqnarray}
\vec k_{12} \equiv {{\epsilon _2\vec p_1 -\epsilon_1 \vec p_2} \over {\epsilon_1+\epsilon_2} },
\label{veck}
\end{eqnarray}
with 
\begin{eqnarray}
\epsilon_i\equiv {1\over 2} (E_i +o_i ),
\label{epsilon}
\end{eqnarray}
and 
\begin{eqnarray}
o_i\equiv \sqrt{m_i^2+k_{12}^2}
\label{omega}
\end{eqnarray}
where we need pay attention that these equations from Eq.(\ref{veck}) to Eq. (\ref{omega}) are 
coupled for $k_{12}$. The momentum $\vec k_{12}$ is regarded an instanteneous 
momentum of center of mass system.

We start to enter the center of mass system, which the boosting velocity $\vec u$ is now chosen instead of $\vec \upsilon$ 
\begin{eqnarray}
\vec u \equiv { {\vec p_1 +\vec p_2 } \over {E_1+E_2}}.
\end{eqnarray}
We have 
\begin{eqnarray}
L \mf{p}_1 = (o_1, \vec k_{12})
\end{eqnarray}
and 
\begin{eqnarray}
L \mf{p}_2 = (o_2, -\vec k_{12}).
\end{eqnarray}
Therefore, we solve them to have $\vec k$ as
\begin{eqnarray}
\vec k_{12}= {1\over 2} \Biggl( (\vec p_1-\vec p_2) 
-(\vec p_1+\vec p_2) \bigl( 
{E_1 -E_2 +{m_1^2-m_2^2 \over \sqrt{(E_1+E_2)^2-(\vec p_1+\vec p_2)^2} } \over
E_1+E_2+\sqrt{(E_1+E_2)^2-(\vec p_1+\vec p_2)^2} }\bigr) \Biggr)  
\end{eqnarray}
We know the following Jacobian ${\cal J}_{12}$ 
\begin{eqnarray}
{\cal J}_{12} \equiv { \partial (\vec p_1 , \vec p_2) \over \partial (\vec k , \vec P) }
={E_1 E_2 \over E_1+E_2 }{o_1+o_2 \over o_1 o_2 }.
\end{eqnarray}
If we take the case of equal mass ($m_1=m_2$), we have $o_1 =o_2$, 
\begin{eqnarray}
\vec k _{12}|_{m_1=m_2} = {1\over 2} \Biggl( (\vec p_1-\vec p_2) 
-(\vec p_1+\vec p_2) \bigl( 
{ E_1 -E_2 
%+{m_1^2-m_2^2 \over \sqrt{(E_1+E_2)^2-(\vec p_1+\vec p_2)^2} } 
\over
E_1+E_2+\sqrt{(E_1+E_2)^2-(\vec p_1+\vec p_2)^2} }\bigr) \Biggr)  
\label{G3.6}
\end{eqnarray}
and 
\begin{eqnarray}
{\cal J}_{12}|_{m_1=m_2} %= { \partial (\vec p_1 , \vec p_2) \over \partial (\vec k , \vec P) }
={E_1 E_2 \over E_1+E_2 }{4 \over \sqrt{(E_1+E_2)^2-(\vec p_1+\vec p_2)^2} } .
\label{G3.16}
\end{eqnarray}
Eq. (\ref{G3.6}) and Eq. (\ref{G3.16}) are corresponding to Eq.(3.6) and (3.16) of \cite{Glockle:1986zz}, respectively. 

\section{Boosted potential} 
\label{Boost_potential}
Let us consider two equal mass particles ($m=m_1=m_2$) which are labeled 1 and 2 in the 2-body center of mass system with interaction $v_{12}$. 
We have  a invariant mass $\sqrt{S_{12}}$ of the system as
\begin{eqnarray}
\sqrt{S_{12}}=2\sqrt{m^2+k_{12}^2}+v_{12}.
\end{eqnarray}
where $k_{12}$ is the relative momentum between particle 1 and 2. 

On the other hand, we leave from the 2-body c. m. system, the total momentum $\vec P_{12}\equiv \vec p_1+\vec p_2$  is nonzero. The invariant mass $\sqrt{S_{12}^{\rm boost} }$ 
is given as 
\begin{eqnarray}
\sqrt{S_{12}^{\rm boost} }= \sqrt {S_{12}+ P_{12}^2}=\sqrt{(2\sqrt{m^2+k_{12}^2}+v_{12})^2 +P_{12}^2}.
\end{eqnarray}
Now, one introduce so-called boosted potential $V_{12}$ as
\begin{eqnarray}
V_{12}(P_{12}) &&\equiv \sqrt{s_{12}^{\rm boost}}- \sqrt{(2\sqrt{m^2+k_{12}^2}+ 0 )^2 +P_{12}^2} \cr && = \sqrt{(2\sqrt{m^2+k_{12}^2}+v_{12})^2 +P_{12}^2} -\sqrt{4(m^2+k_{12}^2)+P_{12}^2}.
\label{V}
\end{eqnarray}
After quantumization ($k_{12} \to \hat k_{12}, v \to \hat v $ and  $V \to \hat V$) the boosted potential operator $\hat V_{12}(P_{12})$ is still a diagonal operator 
to the boosting momentum $P_{12}$. We have a boosted Schr\"odinger equation for the wave function 
$\phi_{12}$ 
\begin{eqnarray}
\big( \sqrt{4(m^2+\hat k_{12}^2)+P_{12}^2} +  \hat V_{12}(P_{12}) \big) \phi_{12} 
= \sqrt{M ^2 + P_{12}^2} \phi_{12}
\end{eqnarray}
and a un-boosted one,
\begin{eqnarray}
\big(2\sqrt{m^2+\hat k_{12}^2} +  \hat v_{12} \big) \phi_{12} = M \phi_{12} .
\end{eqnarray}
where $M$ is a eigen value of mass operator $\sqrt{\hat S_{12}}$.

\section{Relativistic Faddeev Equations}
\label{Relativistic_Faddeev}

For the 3-body system, we add to third equal mass particle. There are 3 piarwises (subsystems) denoted not only as (12) but as (23) and (31).
For each piarwise (ij) we can define the boosted potential 
as Eq. (\ref{V}) by the boosting momentum $\vec P_{ij}$.
\begin{eqnarray}
V_{ij}(P_{ij}) &&\equiv \sqrt{(2\sqrt{m^2+k_{ij}^2}+v_{ij})^2 +P_{ij}^2} -\sqrt{4(m^2+k_{ij}^2)+P_{ij}^2}.
\label{V_ij}
\end{eqnarray}
We choose now the boosting momentum $\vec P_{ij}= - \vec p_k$, which means it is 
3-body c. m. system ($i\ne k\ne j$);
\begin{eqnarray}
\vec P_{ij}=\vec p_i+\vec p_j=-\vec p_k,~~~~ \vec p_1+\vec p_2+\vec p_3=0
\end{eqnarray}
One may naturally have an idea the following 3-body invariant mass  $\sqrt{S_{123}}$ poses a symmetry.
\begin{eqnarray}
\sqrt{S_{123}}&&= \sqrt{m^2+p_1^2}+\sqrt{m^2+p_2^2}+\sqrt{m^2+p_3^2}
+V_{12}(P_{12})+V_{23}(P_{23})+V_{31}(P_{31}) \cr
&&=\sqrt{m^2+p_1^2}+\sqrt{m^2+p_2^2}+\sqrt{m^2+p_3^2}
+V_{12}(p_3)+V_{23}(p_1)+V_{31}(p_2) \cr
&&=\sqrt{(2\sqrt{m^2+k_{12}^2}+v_{12})^2 +p_3^2} +\sqrt{m^2+p_3^2} 
+V_{23}(p_1)+V_{31}(p_2) \cr
&&=\sqrt{(2\sqrt{m^2+k_{23}^2}+v_{23})^2 +p_1^2} +\sqrt{m^2+p_1^2} 
+V_{31}(p_2)+V_{12}(p_3) \cr
&&=\sqrt{(2\sqrt{m^2+k_{31}^2}+v_{31})^2 +p_2^2} +\sqrt{m^2+p_2^2} 
+V_{12}(p_3)+V_{23}(p_1) 
\label{3bodyeq}
\end{eqnarray}
This symmetry helps us to build the relativistic Faddeev equations.
After the quantumization we write the relativistic Faddeev equation for bound state as
\begin{eqnarray}
\phi_{ij} = \hat G_0~\hat t_{ij}~(\phi_{jk} +\phi_{ki}) 
\label{Faddeev}
\end{eqnarray}
where $\phi_{ij}$ is the Faddeev component for the subsystem (ij) % from total wave function $\Psi$, 
\begin{eqnarray}
\phi_{ij} \equiv \hat G_0 \hat V_{ij} \Psi
\label{FC1}
\end{eqnarray}
with the total wave function $\Psi$
\begin{eqnarray}
\Psi=\phi_{12}+\phi_{23}+\phi_{31}
\end{eqnarray}
and $\hat G_0$ is the three-body Green's function, 
\begin{eqnarray}
\hat G_0= {1 \over M_{123} - \bigl( \sqrt{4 (m^2+\hat k_{ij}^2 ) +\hat p_k^2} +\sqrt{m^2+\hat p_k^2} \bigr) }
\end{eqnarray} 
and $ \hat t_{ij}$ is the t-matrix of subsystem (ij) which is satisfactory with the 
LS equation;
\begin{eqnarray}
\hat t_{ij} = \hat V_{ij} + \hat V_{ij}~\hat G_0~\hat t_{ij}.
\end{eqnarray}
where $M_{123}$ is the eigen value of the mass operator $\sqrt{\hat S_{123}}$.

In the case of a system with identical particles, the permutation operators built
from transpositions ${\cal P}_{ij}$, interchanging
particles $i$ and $j$, are used to express all two-body interaction
\begin{eqnarray}
V_{12}+V_{23}+V_{31}\equiv (1+{\cal P}_{12}{\cal P}_{23}+{\cal P}_{13}{\cal P}_{23}) V_{12} = (1+{\cal P}) V,
\end{eqnarray}
where we have singled out the $(12)$ pair and denoted $V\equiv V_{12}$,
$t \equiv t_{12}$ and $\phi_{12} \equiv \phi$ with permutation operator
${\cal P}\equiv {\cal P}_{12}{\cal P}_{23}+{\cal P}_{13}{\cal P}_{23} $.
 We have a simple presentation of Eq. (\ref{Faddeev}).
\begin{eqnarray}
\phi = \hat G_0~\hat t~{\cal P} \phi.
\label{Faddeev1}
\end{eqnarray}

\section{Relativistic Yakubovsky Equations}
\label{Relativistic_Yakubovsky}

For the 4-body system, we add to fourth equal mass particle. There are 6 piarwises (subsystems) denoted not only as (12), (23) and (31) but as (14),(24) and (34).

We choose now the boosting momentum $\vec P_{ij}= - \vec p_k -\vec p_l$, which means it is 
4-body c. m. system ($k\ne i,j,l$, and $l\ne i,j,k$);
\begin{eqnarray}
\vec P_{ij}=\vec p_i+\vec p_j=-\vec p_k -\vec p_l=-\vec P_{kl},~~~~ \vec p_1+\vec p_2+\vec p_3+\vec p_4=0
\end{eqnarray}
One may naturally have an idea the following 4-body invariant mass  $\sqrt{S_{1234}}$ poses a symmetry.
\begin{eqnarray}
\sqrt{S_{1234}}  
%&&= \sqrt{m^2+p_1^2}+\sqrt{m^2+p_2^2}+\sqrt{m^2+p_3^2}+\sqrt{m^2+p_4^2}\cr
%&&+V_{12}(P_{12})+V_{23}(P_{23})+V_{31}(P_{31}) +V_{14}(P_{14})+V_{24}(P_{24})+V_{34}(P_{34}) \cr
&&= \sqrt{m^2+p_1^2}+\sqrt{m^2+p_2^2}+\sqrt{m^2+p_3^2}+\sqrt{m^2+p_4^2}\cr
&&+V_{12}(P_{34})+V_{23}(P_{14})+V_{31}(P_{24}) +V_{14}(P_{23})+V_{24}(P_{31})+V_{34}(P_{12}) %\cr
\label{4bodyeq}
\end{eqnarray}
The most important thing is that during generate the boosting potential $V_{ij}$ the momentum $P_{kl}$ behaves as a parameter. In other word, the boosting potential operator  $\hat V_{ij}$ is diagonal to the momentum $P_{kl}$. 

Using the 3-body relative momentum $\vec q_k$ between the subsystem (ij) and the third particle, and the [2+2] partition relative momentum $\vec s_l$ between the subsystem (ij) and (kl) (see Apppendix \ref{appA}) we rewrite the 4-body invaliant mass $\sqrt{S_{1234}}$ 
\begin{eqnarray}
\sqrt{S_{1234}}=
&&= \sqrt{\bigl(\sqrt{(2\sqrt{m^2+k_{12}^2})^2+q_3^2}+\sqrt{m^2+q_3^2} \bigr)^2+p_4^2}+\sqrt{m^2+p_4^2}\cr
&&+V_{12}^{[3+1]}(q_3;p_4)+V_{23}+V_{31}+V_{14}+V_{24}+V_{34} \cr
&&= \sqrt{\bigl(\sqrt{(2\sqrt{m^2+k_{12}^2}+v_{12})^2+q_3^2}+\sqrt{m^2+q_3^2} \bigr)^2+p_4^2}+\sqrt{m^2+p_4^2}\cr
&&~~~~~~~~~~~~~~~~~~~~~~~~~+V_{23}+V_{31}+V_{14}+V_{24}+V_{34} \cr
&&= \sqrt{(2\sqrt{m^2+k_{12}^2})^2+s_4^2}+\sqrt{(2\sqrt{ m^2+ k_{34}^2 } )^2 + s_4^2 }\cr
&&+V^{[2+2]}_{12}(s_4)+V_{23}+V_{31}+V_{14}+V_{24}+V_{34} \cr
&&= \sqrt{(2\sqrt{m^2+k_{12}^2}+v_{12})^2+s_4^2}+\sqrt{(2\sqrt{ m^2+k_{34}^2 } )^2 + s_4^2 }\cr
&&~~~~~~~~~~~~~~~~~~~~+V_{23}+V_{31}+V_{14}+V_{24}+V_{34}.
\label{4bodyeq2}
\end{eqnarray}
where $q_3$ is a relative momentum between subsystem (12) and the third particle, and $s_4$ is a relative momentum between subsystem (12) and (34). (see Appendix \ref{appA})

The following equations from (\ref{rF}) to (\ref{T3T22}) are simply demonstrated as usual nonrelativistic Yakubovsky form except 
for the relativistic Green's function (\ref{Green4}) and the boosted potential.
Similarily (\ref{FC1}), Faddeev component $\phi_{ij}^{(4)}$ is defined with the four-body total wave function $\Psi^{(4)}$ as
\begin{eqnarray}
\phi_{ij}^{(4)} \equiv \hat G_0^{(4)} \hat  V_{ij} \Psi^{(4)}
\label{rF}
\end{eqnarray}
where $\Psi^{(4)}$ is the total wave function which consists of 6 Faddeev components $\phi_{ij}^{(4)}$
\begin{eqnarray}
\Psi^{(4)} = \sum_{(ij)} \phi_{ij}^{(4)}.
\end{eqnarray}

In case of identical particle, using permutation operator ${\cal P}$, ${\cal P}_{34} $ and $\tilde {\cal P}$ we have Faddeev equations for the four-body system as  
\begin{eqnarray}
\phi^{(4)}=\phi_{12}^{(4)}=
\hat G_0^{(4)} ~ \hat t^{(4)}  ({\cal P}-{\cal P}_{34}{\cal P} + \tilde {\cal P} ) \phi^{(4)}
\equiv (1-{\cal P}_{34}) \psi_1 + \psi_2 
\label{YC}
\end{eqnarray}
with 
\begin{eqnarray}
\tilde {\cal P} \equiv {\cal P}_{13} {\cal P}_{24}
\end{eqnarray}
where we have again singled out the $(12)$ pair and denoted $V\equiv V_{12}$,
$t ^{(4)} \equiv t_{12}^{(4)}$ and $\phi_{12} ^{(4)} \equiv \phi ^{(4)}$,
and  $\hat t ^{(4)} $ also obeys Lippmann-Schwinger equation;
\begin{eqnarray}
\hat t^{(4)} = \hat V + \hat V~\hat G_0^{(4)}~\hat t^{(4)}.
\end{eqnarray}
  and $\hat G_0^{(4)}$ is the four-body Green's function, 
\begin{eqnarray}
\hat G_0^{(4)}= && {1 \over M_{1234} -
\Bigl(\sqrt{\bigl(\sqrt{(2\sqrt{m^2+\hat k_{12}^2})^2+\hat q_3^2}+\sqrt{m^2+\hat q_3^2} \bigr)^2+\hat p_4^2}+\sqrt{m^2+\hat p_4^2} \Bigr) }\cr
&&=
{1 \over M_{1234} -
\Bigl( \sqrt{(2\sqrt{m^2+\hat k_{12}^2})^2+\hat s_4^2}+\sqrt{(2\sqrt{ m^2+ \hat k_{34}^2 } )^2 + \hat s_4^2 } \Bigr) }
\label{Green4}
\end{eqnarray} 
The Yakubovsky components $\psi_1$ and $\psi_2$ 
already appear in Eq.(\ref{YC}) which are defined as
\begin{eqnarray}
\psi_1\equiv \hat G_0^{(4)} ~\hat t ^{(4)} {\cal P} \phi^{(4)} 
\label{YC1}
\end{eqnarray}
\begin{eqnarray}
\psi_2 \equiv \hat G_0^{(4)} ~\hat t ^{(4)} \tilde {\cal P} \phi^{(4)} 
\end{eqnarray}
We show the relativistic Yakubovsky equations for bound state.
\begin{eqnarray}
&&\psi_1=  - \hat G_0^{(4)} ~\hat T{\cal P}_{34} \psi_1 + \hat G_0^{(4)}~\hat T \psi_2, \cr
&&\psi_2=  \hat G_0^{(4)} ~ \hat {\tilde T} \tilde {\cal P} (1-{\cal P}_{34}) \psi_1.
\label{rY}
\end{eqnarray}
where $\hat T$ and $\hat {\tilde T}$ are 3-body t-matrix operator and 2+2 partition t-matrix operator, respectively. 
\begin{eqnarray}
&&\hat T= \hat t^{(4)} + \hat t^{(4)}~{\cal P}~\hat G_0^{(4)}~\hat T,\cr
&&\hat {\tilde T}= \hat t^{(4)} + \hat t^{(4)}~ \tilde {\cal P}~ \hat G_0^{(4)}\hat {\tilde T}.
\label{T3T22}
\end{eqnarray}
The relativistic Yakubovsky equation (\ref{rY}) keeps similar form as nonrelativistic one.

\section{Inclusion of 3-body force}
\label{Inclusion3NF}

Recently \cite{Huber:1996cg,Kamada:2019irm}, we obtained the Faddeev equations and Yakubovsky equations 
including 3-body force.
The 3-body force $w_{123}$ is naturally decomposed into three parts
\begin{eqnarray}
w_{123}\equiv w_{123}^{(1)}+w_{123}^{(2)}+w_{123}^{(3)} = (1+{\cal P}) w_{123}^{(3)} 
=(1+{\cal P}) w
\end{eqnarray}
with  $w = w_{123}^{(3)} $.
Instead of Eq. (\ref{FC1}) the Faddeev component is defined including a part of 3-body force 
\begin{eqnarray}
\phi = \hat G_0 (\hat V +\hat w ) \Psi
\end{eqnarray}
Including 3-body force the Faddeev equation for bound state is rewritten as
\begin{eqnarray}
\phi = \hat G_0 \hat \tau \phi
\label{incF}
\end{eqnarray}
where $\hat \tau$ is defined as
\begin{eqnarray}
\hat \tau \equiv \hat t {\cal P} + (1+\hat t \hat G_0) \hat w(1+{\cal P}).
\end{eqnarray}
Because the 3-body force are given in 3-body center of mass system, we need not boost the 3-body force.

On the other hand, in case of the 4-body system, the 3-body force $\hat w_{123}$ 
is required to be boosted with the direction of the last momentum $\vec p_4$.
The invariant mass of the 4-body system $\sqrt{S_{1234}}$ may be given as 
\begin{eqnarray}
\sqrt{S_{1234}}  
&&= \sqrt{m^2+p_1^2}+\sqrt{m^2+p_2^2}+\sqrt{m^2+p_3^2}+\sqrt{m^2+p_4^2}\cr
&&+V_{12}+V_{23}+V_{31}+V_{14}+V_{24}+V_{34} \cr
&&+W_{123}(p_4)+W_{234}(p_1)+W_{341}(p_2)+W_{412}(p_3)\cr
&&= \sqrt{m^2+p_1^2}+\sqrt{m^2+p_2^2}+\sqrt{m^2+p_3^2}+\sqrt{m^2+p_4^2}\cr
&&+V_{12}+V_{23}+V_{31}+V_{14}+V_{24}+V_{34} \cr
&&+W_{123}^{(1)}(p_4)+W_{234}^{(2)}(p_1)+W_{341}^{(3)}(p_2)+W_{412}^{(4)}(p_3)\cr
&&+W_{123}^{(2)}(p_4)+W_{234}^{(3)}(p_1)+W_{341}^{(4)}(p_2)+W_{412}^{(1)}(p_3)\cr
&&+W_{123}^{(3)}(p_4)+W_{234}^{(4)}(p_1)+W_{341}^{(1)}(p_2)+W_{412}^{(2)}(p_3)\cr
&&= \sqrt{\bigl(\sqrt{4(m^2+k_{12}^2)+q_3^2}+\sqrt{m^2+q_3^2}+w_{123}^{(3)}\bigr)^2+p_4^2}+\sqrt{m^2+p_4^2}\cr
&&+V_{12}+V_{23}+V_{31}+V_{14}+V_{24}+V_{34} \cr
&&+W_{123}^{(1)}(p_4)+W_{234}^{(2)}(p_1)+W_{341}^{(3)}(p_2)+W_{412}^{(4)}(p_3)\cr
&&+W_{123}^{(2)}(p_4)+W_{234}^{(3)}(p_1)+W_{341}^{(4)}(p_2)+W_{412}^{(1)}(p_3)\cr
&&~~~~~~~~~~~~~~~~~+W_{234}^{(4)}(p_1)+W_{341}^{(1)}(p_2)+W_{412}^{(2)}(p_3)
\label{4bodyeq3}
\end{eqnarray}
where $W_{ijk}^{(i)}(p_l)$, $W_{ijk}^{(j)}(p_l)$ and $W_{ijk}^{(k)}(p_l)$ are boosted 3-body forces. (see Appendix \ref{appB})
Similarily, instead of Eq. (\ref{YC1}) the Yakubovsky  component is defined 
including a part of boosted 3-body force $\hat W\equiv \hat W_{123}^{(3)}$,
\begin{eqnarray}
&&\psi_1\equiv \hat G_0^{(4)} ~\hat t ^{(4)} {\cal P} \phi^{(4)} 
+(1+\hat G_0^{(4)} \hat t^{(4)} ) \hat G_0 ^{(4)} \hat W \Psi^{(4)},\cr
&&\psi_2\equiv \hat G_0^{(4)} \hat t ^{(4)} \tilde {\cal P} \phi^{(4)}
\end{eqnarray}
where $\Psi^{(4)}$ is the total wave function 
\begin{eqnarray}
\Psi^{(4)} =( 1+{\cal P} -{\cal P}_{34} {\cal P} +\tilde {\cal P})(\psi_1 -{\cal P}_{34}\psi_1 +\psi_2).
\end{eqnarray}
Yakubovsky 4-body equations for bound state are rewritten as
\begin{eqnarray}
\psi_1=&&  - \hat G_0^{(4)} \hat {\cal T} {\cal P}_{34} \psi_1 +\hat G_0^{(4)} ~\hat {\cal T}~\psi_2 \cr 
&&+(1 + \hat G_0^{(4)}~\hat {\cal T} )(1 + \hat G_0^{(4)}~\hat t^{(4)})~\hat G_0^{(4)} 
\hat W( -{\cal P}_{34} {\cal P} + \tilde {\cal P} ) ( \psi_1 -{\cal P}_{34} ~ \psi_1 +\psi_2), \cr
\psi_2=&& \hat G_0^{(4)} \hat {\tilde T} \tilde {\cal P} (1-{\cal P}_{34}) \psi_1.
\label{rYw3}
\end{eqnarray}
with 
\begin{eqnarray}
\hat {\cal T} = \hat \tau^{(4)} + \hat \tau^{(4)} \hat G_0^{(4)} \hat {\cal T}.
\end{eqnarray}

\section{Conclusion}
\label{Conclusion}
The relativistic Faddeev 3-body equations and the relativistic Yakubovsky 4-body equations are 
given in Eq.(\ref{Faddeev1}) and Eq.(\ref{rY}), respectively. These equations are not deformed from the original ones with 
the boosted potentials in Eq. (\ref{V}) and Eq.(\ref{4bodyeq2}). Inclusion of 3-body force is treated consistently in Faddeev equations (\ref{incF}) and Yakubovsky ones (\ref{rYw3}) 
under Poincar\'e invariance.
These boosted potentials and the boosted 3-body force are
defined by Eqs. (\ref{V[3+1]}), (\ref{V[2+2]}) and (\ref{W}) in Appendices \ref{appA} and \ref{appB}. The quadratic integral equations using the iteration method~\cite{Kamada:2007ms}, in order to obtain these boosted potentials and 3-body force, are demonstrated in Appendix \ref{appC}.

\section*{Acknowledgements}
\label{Acknow}
%Acknowledgements should follow immediately after the conclusion.
Author (H.K.) would like to thank H. Wita\l a, J. Golak, R. Skibi\'nski, K. Topolnicki, A. Nogga and  E. Epelbaum for 
fruitful discussions during the 4th LENPIC meeting in Bochum, Germany.  website \url{http://www.lenpic.org/}
%~\cite{LENPIC:2019lp}. 

% TODO: include author contributions
%\paragraph{Author contributions}
%This is optional. If desired, contributions should be succinctly described in a single short paragraph, using author initials.

% TODO: include funding information
%\paragraph{Funding information}
%Authors are required to provide funding information, including relevant agencies and grant numbers with linked author's initials. Correctly-provided data will be linked to funders listed in the \href{https://www.crossref.org/services/funder-registry/}{\sf Fundref registry}.

\begin{appendix}

\section{Appendix A}
\label{appA}
%Add material which is better left outside the main text in a series of Appendices labeled by capital letters.
 The relative momentum $\vec q_3$ between subsystem (12) and the third particle is given as
\begin{eqnarray}
\vec q_3 \equiv {\epsilon_{12} \vec p_3 -\epsilon_{3} \vec P_{12}  \over \epsilon_{12} +\epsilon_{3}}
\end{eqnarray}
with 
\begin{eqnarray}
\epsilon_{12}={1\over 2} \bigl(E_1+E_2+ \sqrt{(2\sqrt{m^2+k_{12}^2} )^2+ q_3^2} \bigr).
\end{eqnarray}
Using the momentum $\vec q_3$ we have the boosted potential  $V_{12}^{[3+1]} (q_3;p_4)$ in Eq. (\ref{4bodyeq2}).
\begin{eqnarray}
V_{12}^{[3+1]}(q_3;p_4) \equiv &&  \sqrt{\bigl(\sqrt{(2\sqrt{m^2+k_{12}^2}+v_{12})^2+q_3^2}+\sqrt{m^2+q_3^2} \bigr)^2+p_4^2} \cr 
&&- \sqrt{\bigl(\sqrt{(2\sqrt{m^2+k_{12}^2})^2+q_3^2}+\sqrt{m^2+q_3^2} \bigr)^2+p_4^2}.
\label{V[3+1]}
\end{eqnarray}

%\section{Appendix B}
%\label{appB}

The [2+2] partition relative momentum $\vec s_4$ between subsystem (12) and (34) is given as
\begin{eqnarray}
\vec s_4 \equiv {\epsilon_{12,34} \vec P_{34}  -\epsilon_{34,12} \vec P_{12}  \over \epsilon_{12,34} +\epsilon_{34,12}}
\end{eqnarray}
with 
\begin{eqnarray}
&&\epsilon_{12,34}={1\over 2} \bigl(E_1+E_2 + \sqrt{(2\sqrt{m^2+k_{12}^2} )^2+ s_4^2} \bigr),\cr
&&\epsilon_{34,12}={1\over 2} \bigl(E_3+E_4 + \sqrt{(2\sqrt{m^2+k_{34}^2} )^2+ s_4^2} \bigr).
\end{eqnarray}
Using the momentum $\vec s_4$ we have the boosted potential  $V_{12}^{[2+2]} (s_4)$ in Eq. (\ref{4bodyeq2}).
\begin{eqnarray}
V_{12}^{[2+2]}(s_4) \equiv && 
 \sqrt{(2\sqrt{m^2+k_{12}^2}+v_{12})^2+s_4^2} -  \sqrt{(2\sqrt{m^2+k_{12}^2})^2+s_4^2} = V_{12} (s_4)
\label{V[2+2]}
 \end{eqnarray}
Actually, this is the same definition of  $V_{12}$ in Eq. (\ref{V}).

\section{Appendix B}
\label{appB}
The boosted 3-body force $W_{123}^{(3)}(p_4)$ is defined as
\begin{eqnarray}
W_{123}^{(3)}(p_4)\equiv && \sqrt{\bigl(\sqrt{(2\sqrt{m^2+k_{12}^2})^2+q_3^2}+\sqrt{m^2+q_3^2} +w_{123}^{(3)} \bigr)^2+p_4^2} \cr
&& -\sqrt{\bigl(\sqrt{(2\sqrt{m^2+k_{12}^2})^2+q_3^2}+\sqrt{m^2+q_3^2} \bigr)^2+p_4^2}, \cr
W_{123}^{(\eta)}(p_4)\equiv && \sqrt{\bigl(\sqrt{(2\sqrt{m^2+k_{12}^2})^2+q_3^2}+\sqrt{m^2+q_3^2} +w_{123}^{(\eta)} \bigr)^2+p_4^2} \cr
&& -\sqrt{\bigl(\sqrt{(2\sqrt{m^2+k_{12}^2})^2+q_3^2}+\sqrt{m^2+q_3^2} \bigr)^2+p_4^2}.
\label{W}
\end{eqnarray}
where $\eta$ can be chosen $1,2$ or $3$.

\section{Appendix C}
\label{appC}

The boosted potential $V_{12} (\vec k,\vec k';q) \equiv \langle \vec k |\hat V_{12} (q) | \vec k' \rangle$ of
Eq.(\ref{V})
is a solution of the following quadratic integral equation~\cite{Kamada:2007ms}.
\begin{eqnarray}
V_{12}(\vec k,\vec k';q)= && \displaystyle \frac{1
%\sqrt{4(m^{2}+k^{2})}+\sqrt{4(m^{2}+k^{\prime 2})}
}{\sqrt{4(m^{2}+k^{2})+q^{2}}+\sqrt{4(m^{2}+k^{\prime 2})+q^{2}}} \cr 
&& \times \Biggl( \bigl( \sqrt{4(m^{2}+k^{2})}+\sqrt{4(m^{2}+k^{\prime 2})} \bigr)~v_{12}(\vec k,\vec k') \cr
&&~~~~+\displaystyle \int\left(v_{12}(\vec k,\vec k^{''})v_{12}(\vec k^{''},\vec k')-V_{12}(\vec k,\vec k'';q)V_{12}(\vec k'',\vec k';q)\right)d^{3}k'' \Biggr)
\label{eqV}
\end{eqnarray}
with 
\begin{eqnarray}
v_{12}(\vec k,\vec k') \equiv \langle \vec k | \hat v_{12} | \vec k' \rangle .
\end{eqnarray}
The boosted potential $V_{12}^{[3+1]} (\vec k,\vec k';q,r) \equiv 
\langle \vec k |\hat V_{12}^{[3+1]} (q;r) | \vec k' \rangle$ of
Eq.(\ref{V[3+1]})
is a solution of the following quadratic integral equation.
\begin{eqnarray}
&&V_{12}^{[3+1]}(\vec k,\vec k';q,r) \cr
&&=\displaystyle \frac{
%\bigl(
%\sqrt{4(m^{2}+k^{2})+q^{2}}+\sqrt{4(m^{2}+k^{\prime 2})+q^{2}}+2\sqrt{m^{2}+q^{2}} \bigr) 
%V_{12}(\vec k,\vec k';q)
1}{\sqrt{\left(\sqrt{4(m^{2}+k^{2})+q^{2}}+\sqrt{m^{2}+q^{2}}\right)^{2}+r^{2}}+\sqrt{\left(\sqrt{4(m^{2}+k^{\prime 2})+q^{2}}+\sqrt{m^{2}+q^{2}}\right)^{2}+r^{2}}} \cr
&&\times \Biggl( \bigl( \sqrt{4(m^{2}+k^{2})+q^{2}}+\sqrt{4(m^{2}+k^{\prime 2})+q^{2}}+2\sqrt{m^{2}+q^{2}} \bigr) ~V_{12}(\vec k,\vec k';q) \cr
&&
~~~~+\displaystyle \int\left(V_{12}(\vec k,\vec k'';q)V_{12}(\vec k'',\vec k';q)-V_{12}^{[3+1]}(\vec k,\vec k'';q,r)V_{12}^{[3+1]}(\vec k'',\vec k';q,r)\right)d^{3}k'' \Biggr).
\end{eqnarray}
Finally, the boosted 3-body force $W_{123}^{(3)} (\vec k,\vec q,\vec k',\vec q' ;r) 
\equiv \langle \vec k  \vec q |\hat W_{123}^{(3)} (r) | \vec k' \vec q' \rangle$ of
Eq.(\ref{W})
is a solution of the following quadratic integral equation.
\begin{eqnarray}
&&W_{123}^{(3)}(\vec k,\vec q,\vec k',\vec q';r)\cr
&&=\displaystyle \frac{1
%\bigl( \sqrt{4(m^{2}+k^{2})+q^{2}}+\sqrt{4(m^{2}+k^{\prime 2})+q^{\prime 2}}+\sqrt{m^{2}+q^{2}}+
%\sqrt{m^{2}+q^{\prime 2}} \bigr) w_{123}^{(3)}(\vec k,\vec q,\vec k',\vec q')
}{\sqrt{\left(\sqrt{4(m^{2}+k^{2})+q^{2}}+\sqrt{m^{2}+q^{2}}\right)^{2}+r^{2}}+\sqrt{\left(\sqrt{4(m^{2}+k^{\prime 2})+q^{\prime 2}}+\sqrt{m^{2}+q^{\prime 2}}\right)^{2}+r^{2}}}
%w_{123}^{(3)}(k,q,k',q')$
\cr
&&\times \Biggl( \bigl( \sqrt{4(m^{2}+k^{2})+q^{2}}+\sqrt{4(m^{2}+k^{\prime 2})+q^{\prime 2}}+\sqrt{m^{2}+q^{2}}+
\sqrt{m^{2}+q^{\prime 2}} \bigr) w_{123}^{(3)}(\vec k,\vec q,\vec k',\vec q') \cr
&&+\displaystyle \int\int\left(w_{123}^{(3)}(\vec k,\vec q,\vec k'',\vec q'')w_{123}^{(3)}(\vec k'',\vec q'',\vec k',\vec q') 
-W_{123}^{(3)}(\vec k,\vec q,\vec k'',\vec q'';r)W_{123}^{(3)}(\vec k'',\vec q'',\vec k',\vec q';r)\right)\cr
&& \times d^{3}k'' d^{3}q'' \Biggr)   %\cr &&
\label{Weq}
\end{eqnarray}
with 
\begin{eqnarray}
w_{123}^{(3)}(\vec k,\vec q, \vec k',\vec q') 
\equiv \langle \vec k \vec q | \hat w_{123}^{(3)} | \vec k' \vec q' \rangle .
\end{eqnarray}

These equations from (\ref{eqV}) to (\ref{Weq}) may be solved by the iteration method \cite{Kamada:2007ms}.

\end{appendix}

% TODO:
% Provide your bibliography here. You have two options:

% FIRST OPTION - write your entries here directly, following the example below, including Author(s), Title, Journal Ref. with year in parentheses at the end, followed by the DOI number.
%\begin{thebibliography}{99}
%\bibitem{1931_Bethe_ZP_71} H. A. Bethe, {\it Zur Theorie der Metalle. i. Eigenwerte und Eigenfunktionen der linearen Atomkette}, Zeit. f{\"u}r Phys. {\bf 71}, 205 (1931), \doi{10.1007\%2FBF01341708}.
%\bibitem{arXiv:1108.2700} P. Ginsparg, {\it It was twenty years ago today... }, \url{http://arxiv.org/abs/1108.2700}.
%\end{thebibliography}

% \bibliographystyle{SciPost_bibstyle} % Include this style file here only if you are not using our template

%\bibliography{Kamada_1.bib}

\nolinenumbers

\end{document}